\documentclass[10pt,conference]{IEEEtran}
\usepackage{breakcites}
\usepackage{balance}
\usepackage{graphicx} 
\usepackage[tight,footnotesize]{subfigure}
\usepackage{caption,setspace}
\usepackage{epstopdf}
\usepackage{bbm}
\usepackage{tabularx, ragged2e}
\usepackage[tight,footnotesize]{subfigure}
\usepackage{mathtools,amssymb,lipsum}
\usepackage{cuted}
\usepackage{lipsum}
\usepackage{mathtools}
\usepackage{cuted}
\usepackage{xcolor}
\usepackage{amsthm}
\usepackage{stfloats}
\usepackage{soul}
\usepackage{amsmath,amssymb}

\usepackage[nolist]{acronym}
\usepackage[normalem]{ulem}

\usepackage[utf8]{inputenc}
\usepackage[english]{babel}
\usepackage[utf8]{inputenc}
\usepackage{amsmath}
\usepackage{amsfonts}
\usepackage{graphicx}
\usepackage[colorinlistoftodos]{todonotes}
\usepackage[ruled,vlined]{algorithm2e}
\usepackage{algpseudocode}
\usepackage[super]{nth}
\usepackage{comment}

\newcommand{\red}[1] {\textcolor[rgb]{0.0,0.0,0.0}{{#1}}}

\begin{acronym} 
\acro{3GPP}{third generation partnership project}
\acro{2D}{two-dimensional}
\acro{2G}{second generation}
\acro{3G}{third generation}
\acro{4G}{fourth generation}
\acro{5G}{fifth generation}
\acro{5GAA}{5G Automotive Association}
\acro{5GS}{5G system}
\acro{ADAS}{advanced driver assistance systems}
\acro{AF}{application function}
\acro{AMC}{adaptive modulation and coding}
\acro{AS}{application server}
\acro{AWGN}{additive white Gaussian noise}
\acro{BEP}{beacon error probability}
\acro{BM-SC}{Broadcast Multicast Service Center}
\acro{BP}{beacon periodicity}
\acro{B-CSA}{broadcast \ac{CSA}}
\acro{BSM}{basic safety message}
\acro{ICW}{intersection collision warning}
\acro{BSS}{basic service set}
\acro{C-ITS}{cooperative-intelligent transport systems}
\acro{C-NOMA}{Cooperative \ac{NOMA}}
\acro{C-V2X}{cellular-\ac{V2X}}
\acro{CA}{collision avoidance}
\acro{CAM}{cooperative awareness message}
\acro{CAV}{connected and autonomous vehicle}
\acro{CBR}{channel busy ratio}
\acro{cdf}{cumulative distribution function}
\acro{CDMA}{code-division multiple access}
\acro{CoMP}{coordinated multi-point}
\acro{CN}{core network}
\acro{CP-OFDM}{cyclic prefix \ac{OFDM}}
\acro{CPM}{collective perception message}
\acro{CR}{channel occupancy ratio}
\acro{CRDSA}{contention resolution diversity slotted ALOHA}
\acro{CSA}{coded-slotted ALOHA}
\acro{CSD}{cyclic shift diversity}
\acro{CSI}{channel state information}
\acro{CSIT}{channel state information at the transmitter}
\acro{CSIR}{channel state information at the receiver}
\acro{CSMA/CA}{carrier sense multiple access with collision avoidance}
\acro{D2D}{device-to-device}
\acro{DCC}{decentralized congestion control}
\acro{DCF}{distributed coordination function}
\acro{DCI}{downlink control
information}
\acro{DCM}{Dual carrier modulation}
\acro{DENM}{decentralized environmental notification message}
\acro{DMRS}{demodulation reference signal}
\acro{DRX}{discontinuous reception}
\acro{DSRC}{dedicated short range communication}
\acro{DOT}{Department of Transportation}
\acro{DTU}{device under test}
\acro{EC}{European Commission}
\acro{ECP}{extended cyclic prefix}
\acro{EDCA}{enhanced distributed coordination access}
\acro{EDGE}{Enhanced Data rates for GSM Evolution}
\acro{eMBMS}{evolved Multicast Broadcast Multimedia Service}
\acro{eMMB}{enhanced mobile broadband}
\acro{EMU}{external management unit}
\acro{eNodeB}{evolved NodeB}
\acro{EPC}{Evolved Packet Core}
\acro{EPS}{Evolved Packet System}
\acro{ETSI}{European telecommunications standards institute}
\acro{EVW}{emergency vehicle warning}
\acro{FCC}{Federal Communications Commission}
\acro{FCD}{floating car data}
\acro{FD}{in-band full-duplex}
\acro{FDD}{frequency division duplex}
\acro{FDMA}{frequency division multiple access}
\acro{FEC}{forward error correction}
\acro{GNSS}{global navigation satellite system}
\acro{GPRS}{general packet radio service }
\acro{GPS}{global positioning system}
\acro{HARQ}{hybrid automatic repeat request}
\acro{HD}{half-duplex}
\acro{HiL}{hardware-in-the-loop}
\acro{HCU}{hybrid control unit}
\acro{HSDPA}{High Speed Downlink Packet Access}
\acro{HSPA}{High Speed Packet Access}
\acro{IBE}{in-band emission}
\acro{ICI}{Inter-carrier interference}
\acro{IDMA}{interleave-division multiple access}
\acro{IM}{index modulation}
\acro{IP}{Internet Protocol}
\acro{IPG}{inter-packet gap}
\acro{IR}{incremental redundancy}
\acro{ISI}{inter-symbol interference}
\acro{ITS}{intelligent transport system}
\acro{ITS-S}{ITS-station}
\acro{i.i.d.}{independent identically distributed}
\acro{KPI}{key performance indicator}  
\acro{LDPC}{low-density parity-check}
\acro{LDS}{low-density spreading}
\acro{LIDAR}{light detection and ranging}
\acro{LOS}{line-of-sight}
\acro{LTE}{long term evolution}  
\acro{LTE-D2D}{\ac{LTE} with \ac{D2D} communications}  
\acro{LTE-V2X}{long-term-evolution-vehicle-to-anything} \acro{LTE-V2V}{\ac{LTE}-\ac{V2V}}
\acro{MAC}{medium access control}
\acro{MBMS}{Multicast Broadcast Multimedia Service}
\acro{MBMS-GW}{MBMS Gateway}
\acro{MBSFN}{MBMS Single Frequency Network}
\acro{MCE}{Multi-cell Coordination Entity}
\acro{MEC}{mobile edge computing}
\acro{MCM}{maneuver coordination message}
\acro{MCS}{modulation and coding scheme}
\acro{MIMO}{multiple input multiple output}
\acro{MRD}{maximum reuse distance}
\acro{MUD}{multi-user detection}
\acro{NEF}{network exposure function}
\acro{NGV}{Next Generation V2X}
\acro{NHTSA}{National Highway Traffic Safety Administration}
\acro{NLOS}{non-line-of-sight}
\acro{NLOSb}{non-line-of-sight due to buildings}
\acro{NLOSv}{non-line-of-sight due to vehicles}
\acro{NOMA}{non-orthogonal multiple access}
\acro{NOMA-MCD}{\ac{NOMA}-mixed centralized/distributed}
\acro{NR}{new radio}
\acro{OBU}{on-board unit}
\acro{OCB}{outside of the context of a basic service set}
\acro{OEM}{Original equipment manufacturer}
\acro{OFDM}{orthogonal frequency-division multiplexing}
\acro{OFDMA}{orthogonal frequency-division multiple access}
\acro{OMA}{orthogonal multiple access}
\acro{OTFS}{orthogonal time frequency space}
\acro{pdf}{probability density function}
\acro{PAPR}{peak to average power ratio}
\acro{PCF}{policy control function}
\acro{PCM}{platoon control message}
\acro{PD}{packet delay}
\acro{PEP}{pairwise error probability} 
\acro{PER}{packet error rate}
\acro{PIAT}{packet inter-arrival time}
\acro{PSCCH}{Physical Sidelink Control Channel}
\acro{PSFCH}{Physical Sidelink Feedback channel}
\acro{PSSCH}{Physical Sidelink Shared channel}
\acro{PHY}{physical}
\acro{PL}{path loss}
\acro{PLMN}{Public Land Mobile Network}
\acro{PPPP}{ProSe Per-Packet Priority}
\acro{PPPR}{ProSe Per-Packet Reliability}
\acro{ProSe}{Proximity-based Services}
\acro{PRR}{packet reception ratio}
\acro{QAM}{quadrature amplitude modulation}
\acro{QC-LDPC}{quasi-cyclic \ac{LDPC}}
\acro{QoS}{quality of service}
\acro{QPSK}{quadrature phase shift keying}
\acro{RAN}{Radio Access Network}
\acro{RAT}{radio access technology}
\acro{RB}{resource block}
\acro{RBP}{resource block pair}
\acro{RF}{radio frequency}
\acro{RIS}{re-configurable intelligent  surface}
\acro{RR}{radio resource}
\acro{RRC}{radio resource control}
\acro{RSRP}{reference signal received power}
\acro{RSSI}{Received Signal Strength Indicator}
\acro{RSU}{road side unit} 
\acro{SAI}{Service Area Identifier}
\acro{SC-FDMA}{single carrier frequency division multiple access}
\acro{SC-PTM}{Single Cell Point To Multipoint}
\acro{SCS}{subcarrier spacing}
\acro{SCMA}{sparse-code multiple access}
\acro{SCI}{sidelink control information}
\acro{SDN}{Software-defined networking}
\acro{SFFT}{symplectic finite Fourier transform}
\acro{SHINE}{simulation platform for heterogeneous interworking networks}
\acro{SI}{self-interference}
\acro{SIC}{successive interference cancellation}
\acro{SINR}{signal-to-interference-plus-noise ratio}
\acro{SLN}{speed limits notification}
\acro{S-UE}{scheduling UE}
\acro{SNR}{signal-to-noise ratio}
\acro{SPS}{semi-persistent scheduling}
\acro{SRS}{sounding reference signal}
\acro{STBC}{space time block codes}
\acro{SUMO}{simulation of urban mobility}
\acro{SV}{smart vehicle}
\acro{SVI} {slow vehicle indication}
\acro{TB}{transport block}
\acro{TBC}{time before change}
\acro{TBE}{time before evaluation}
\acro{TCU}{telecommunication control unit}
\acro{TDD}{time-division duplex}
\acro{TDMA}{time-division multiple access}
\acro{TTI}{transmission time interval}
\acro{TRS}{scenario/traffic simulator}
\acro{TraCI}{traffic control interface}
\acro{UAV}{unmanned aerial vehicle}
\acro{UD}{Update delay}
\acro{UMTS}{universal mobile telecommunications system}
\acro{URLLC}{ultra-reliable and ultra-low latency communications}
\acro{USIM}{Universal Subscriber Identity Module}
\acro{UTDOA}{uplink time difference of arrival}
\acro{VCG}{vehicle central gateway}
\acro{V2C}{vehicle-to-cellular} 
\acro{V2N}{vehicle-to-network}
\acro{V2I}{vehicle-to-infrastructure} 
\acro{V2P}{vehicle-to-pedestrian}
\acro{V2R}{vehicle-to-roadside}
\acro{V2V}{vehicle-to-vehicle} 
\acro{V2X}{vehicle-to-everything}
\acro{VAM}{vulnerable road user awareness message}
\acro{VRU}{vulnerable road user}
\acro{VUE}{vehicular user equipment}
\acro{WAVE}{wireless access in vehicular environment}
\acro{WBSP}{wireless blind spot probability}
\end{acronym}

\begin{document}

	\title{Optimizations for Hardware-in-the-Loop-Based V2X Validation Platforms}
	
	\author{
		\IEEEauthorblockN{ Babak Mafakheri$^1$, Pierpaolo Gonnella$^2$, Alessandro Bazzi$^1$,}
		\IEEEauthorblockN{ Barbara Mav\`i Masini$^3$, Michele Caggiano$^2$, Roberto Verdone$^1$}
		
		\IEEEauthorblockA{$^1$~University of Bologna,~Italy~
			Email:~\{babak.mafakheri2,~alessandro.bazzi,~roberto.verdone\}@unibo.it}
		\IEEEauthorblockA{$^2$~FEV Italia,~Italy~
			Email:~\{gonnella,~caggiano\}@fev.com}
		\IEEEauthorblockA{$^3$~CNR-IEIIT,~Italy~
			Email:~barbara.masini@ieiit.cnr.it}
	}

	\maketitle
	
	\begin{abstract}
	Connectivity and automation are increasingly getting importance in the automotive industry, which is observing a radical change from vehicles driven by humans to fully automated and remotely controlled ones. The test and validation of all the related devices and applications is thus becoming a crucial aspect; this is raising the interest on hardware-in-the-loop (HiL) platforms which reduce the need for complicated field trials, thus limiting the costs and delay added to the process. With reference to the test and validation of vehicle-to-everything (V2X) communications aspects, and assuming either sidelink LTE/5G-V2X or IEEE 802.11p/bd technologies, in this work we focus on the real-time HiL simulation of the information exchanged by one vehicle under test and the surrounding, simulated, objects. Such exchange must be reproduced in a time-efficient manner, with elaborations done fast enough to allow testing the applications in real-time. \red{More precisely, we discuss the simulation of non-ideal positioning and channel propagation taking into account current impairments. We also provide details on optimization solutions that allowed us to trade-off minor loss in accuracy with a significant reduction of the computation time burden, reaching up to more than one order of magnitude speed increase in our experiments}. 
	\end{abstract}

	\begin{IEEEkeywords}
		Connected and automated vehicles; Real-time simulation; Hardware-in-the-loop; IEEE 802.11p; Cellular-V2X
	\end{IEEEkeywords}
	
	\IEEEpeerreviewmaketitle

	\section{Introduction}

    By the rapid growth of urbanization worldwide, \acp{CAV} will increasingly play an important role to improve safety in our cities. 
    The promises are to reduce the number of accidents and fatalities, to improve traffic and energy efficiency \cite{wegener2020energy}, to enable commercial applications such as toll collection, and so on \cite{zhou2020evolutionary}. Many applications are indeed being designed and developed for the \ac{ITS}, which are based on \ac{V2X} communications
    \cite{menarini2019trudi, chen2015cooperative, di2019design}. These applications take advantage of information about the status and movements of nearby vehicles and objects to take prompt and smart decisions. To this aim, several messages have been defined both in Europe and in the US, including \acp{BSM}, \acp{CAM}, \acp{DENM}, or are under definition, such as \acp{CPM} or \acp{VAM}. 
    The main wireless technologies that are currently used or expected to be used in short range to exchange these messages are IEEE~802.11p (corresponding to \ac{ITS}-G5 in Europe) and \ac{C-V2X}.
    \footnote{Normally included under the name of \ac{C-V2X} there are the legacy LTE based on the Uu-interface and applied to the \ac{ITS}, sidelink LTE-V2X based on PC5, and also the respective 5G solutions.} 
    Safety is one of the most important aspects of such systems, and conducting extensive tests for the validation and reliability of the applications before moving to production is crucial. Noticeably, performing real-world field tests is very challenging for many reasons; first, providing a controlled environment to perform tests in the desired scenarios imposes relevant extra costs. Second, even assuming large budgets, it is technically not feasible to involve large number of nodes or to address a large number of scenarios. Finally, performing tests on the road implies a significant overhead in terms of manpower and dedicated time. . 
    These types of challenges rise the importance of using simulations and emulations platforms as the initial stage of the validation process. 
    In particular, simulators are often used as first step which come, however, with some level of abstractions such as modeling of radio propagation, \ac{GNSS} accuracy, traffic flow, etc. 
    Besides the fact that not all of the aspects are implemented with the same level of details, simulators can hardly address all the issues that are related to the use of the real hardware and software. 
   
    For these reasons, the use of the so-called \ac{HiL} platforms is raising increasing importance in the validation process of modern cars \cite{eisenbarth2020toward}. Differently from standard simulators, in the design and development of a \ac{HiL} platform the ability to guarantee real-time processing is a critical aspect. Focusing on wireless communications, in most works the challenges of real-world tests are addressed with simulators/emulators  (e.g.,  \cite{gupta2012bsm, fallah2019efficient}) but the solutions do not appear to be able to work at real-time because of the high latency that is introduced. In other cases (e.g., in \cite{shah2019real}), the focus is indeed on real-time simulations but the radio-link communication conditions, which introduce significant delay to the simulators, is not discussed. \red{Moreover, although there are other \ac{HiL} platforms implemented and introduced in the literature, they mostly rely on the traffic simulators to receive the status of the objects; it means that they lack the real effects that may happen due to the GPS inaccuracy or real communication link condition between the objects. For example, in \cite{ma2018hardware-1, ma2018hardwar-2, zulkefli2017hardware} the \ac{HiL} platforms are implemented for different use cases of \acp{CAV}. In \cite{ma2018hardware-1}, only \ac{V2I} communications are considered and 
    in \cite{ma2018hardwar-2}, big attention is paid to the driver assistance system and communication is performed \ac{V2V} only, whereas in \cite{zulkefli2017hardware}, an accurate tracking of target vehicle speed and engine operating points is provided.
    However, none of these works target a mechanism that can guarantee the validity of the received packets by considering the various channel conditions or \ac{GNSS} errors in real time.} 
    
    \red{The reason why the channel quality is normally not considered is that it adds  processing delay hardly compatible with the real time nature of an \ac{HiL} platform. A solution might be to use probabilistic models, such as proposed \cite{boban2016modeling}, which however cannot reproduce a specific environment. Even in \cite{9135714}, where channel quality is indeed considered based on the nodes distance, the links are either assumed all in \ac{LOS} or in \ac{NLOS}.}
    \red{As explained, all the approaches currently proposed in literature have some limitations if used in the validation process of real hardware and software for \acp{CAV} and this motivated us to identify the new solutions proposed in this work.}
    
    \red{The platform under design focuses on a specific vehicle with connectivity and \ac{ADAS}, hereafter called ego-vehicle (EGO), and this paper especially focuses on the software component within the HiL platform which aims at defining} the propagation and positioning aspects \red{affecting the perception of the surrounding at the EGO. The contribution of the paper can be summarized as follows:}
   
    first, we describe the architecture aspects of our validation platform which are relevant for connectivity; second, 
    we introduce a method for reproducing \ac{GNSS} positioning inaccuracy in a controlled environment; then, we discuss the optimization of path-loss evaluations in order to fulfil the stringent timing requirements in large scenarios. 
    Finally, we elaborate on the accuracy of the proposed approach by showing results in a case-study scenario.

\begin{figure}[t]
		\centering
		\includegraphics[scale=0.3,keepaspectratio]{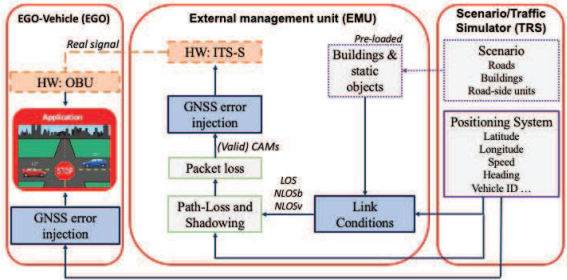}		
		\caption{System Architecture}
		\label{fig:emu}
	\end{figure}

	\section{System Architecture} \label{sec:Model}
	\ac{HiL} validation platforms are typically constituted by hardware parts such as connected \acp{RSU} and \acp{OBU}, and software parts such as channel simulators or road traffic simulators; in our case, they are jointly cooperating with \ac{ADAS} components to evaluate and validate the performance of CAV implementations. 
	The system architecture of the platform under development related to connectivity is represented in Fig.~\ref{fig:emu}. It is composed of three main components (from right to left in the figure): (i) the simulation of the environment and vehicle movements, performed by the \ac{TRS}; (ii) the generation, through the \ac{EMU}, of messages produced by the vehicles and other nodes around the device under test, i.e., the \red{EGO}; and (iii) 
	the \red{EGO} itself, where the applications to be validated are implemented.
	
	On the one hand, the \ac{TRS} is dedicated to the simulation of the road layout, the mobility of all the vehicles and possible pedestrians or bikes, and the buildings. In the results shown later, as \ac{TRS} we have adopted the open-source \ac{SUMO} \cite{SUMO2018} that provides outputs to both the \ac{EMU} (building and device positions) and the EGO (\red{EGO} position).  
	On the other hand, the \ac{EMU} receives the position of the buildings and other static objects at the beginning of the simulation, and an updated position of the moving nodes during the simulation from the \ac{TRS}. Accordingly, it evaluates the quality of the links and determines which and when messages are to be transmitted from the surrounding nodes to the \red{EGO}. The process goes through the identification of propagation conditions (further elaborated in Section~\ref{sec:Method}), the calculation of the received power, the evaluation of lost messages, the addition of realistic GNSS inaccuracies (further elaborated in Section~\ref{sec:GNSS}), and the transmission through a real \ac{ITS-S}, which can be either an \ac{OBU} or an \ac{RSU}. 
	Finally, the EGO receives its own simulated positions from the \ac{TRS} (possibly alters them to account the realistic GNSS inaccuracies) and the information from the neighboring devices from the \ac{EMU}, through the hardware \ac{OBU}. All the information is processed by the applications installed on the \red{EGO} and the correct operation is evaluated.
	
    At the time of writing this paper, virtual \acp{OBU} and \acp{RSU} are being used, emulating both IEEE~802.11p and side-link LTE-V2X communications \cite{bazzi2020wireless}.\footnote{Both IEEE~802.11p-based and sidelink LTE-V2X OBUs and RSUs, produced by Cohda Wireless, are under test in our laboratories and their integration will be performed in the coming weeks.} It is remarkable that the focus of this paper is the optimization of parts that will remain via software, as detailed in the following sections, in order to allow real-time processing, and thus the use of virtual devices do not alter the validity of the detailed solutions. Moreover, in the initial implementation of the platform, the focus is on scenarios with a limited number of vehicles around the \red{EGO} and thus a low level of channel usage. For this reason, the impact of collisions is initially neglected and will be added as a second step.

	\section{Simulating localization inaccuracies}\label{sec:GNSS}
	
	One aspect that needs careful consideration in the validation process of V2X applications is the impact of localization. In fact, most of the applications rely on information about the position of the \red{EGO} and other objects. In the real implementation, the latitude, longitude, and height of the \red{EGO} are derived from the on-board \ac{GNSS} receiver, while the position of the other devices is embedded in the messages received from the other nodes, which are on their own generated on the basis of information obtained from \ac{GNSS} receivers. In the platform, the exact position is provided by the \ac{TRS}. 
	
	The positioning systems do provide the location with some degree of inaccuracy. 
	To take into account the position inaccuracies in our platform, a random variable error is introduced to the exact position of the vehicles provided by \ac{TRS}, both inside the EGO and the EMU.

\begin{figure}[]
		\centering
		\includegraphics[scale=0.23,keepaspectratio]{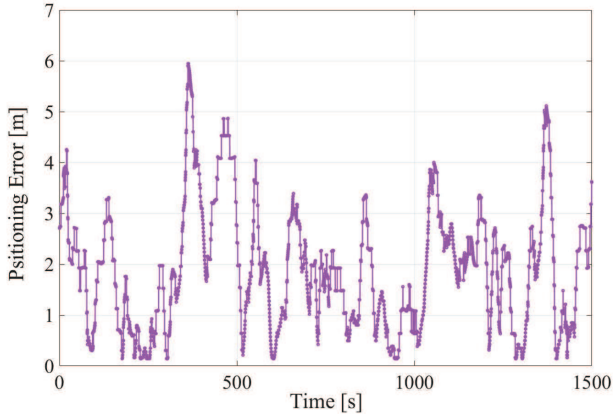}
		\caption{\red{Positioning error vs. time variation.}}
		\label{fig:gps-test}
	\end{figure}

	The literature about localization accuracy is quite fragmented and, at the best of the authors' knowledge, a complete and unified model to represent the error accuracy is not available \cite{bais2016analysis}, \cite{bierman1995error}. 
	Hence, to evaluate the reliability of the \ac{GNSS}, we performed experiments on the road with a \ac{GPS} device. 
	In particular, we positioned the car in a fixed location in the city of Bologna (Italy), collected the information received by the \ac{ITS-S}, and elaborated the outline. 
	As an example of the performed elaboration, Fig.~\ref{fig:gps-test} shows the positioning error in time compared to the long-term average assumed as the ground-truth. As visible, the error  varies significantly during the time, with peaks of more than 5~m. It is also observable that the error is highly correlated, which is an expected effect of the relative movement between the satellites and the Earth.
	In addition, we also performed experiments traveling several times along a given route in order to verify the validity of our conclusions also while moving on the road.
	 
	From the described measurements, we derived a positioning error with an absolute value following a Gaussian distribution with zero mean and a given standard deviation and an angle distributed uniformly between $0$ and $2\pi$. The distance root mean square error obtained in our measurements was equal to 2.32~m. 
	
	In order to reproduce a correlated error, the model proposed for the shadowing in \cite{3GPPTR36.885V14.0.0} was used as a reference. 
	In particular, the correlated magnitude, denoted as $\mu$, and angle, denoted as $\theta$, are derived from the following equations
	\begin{flalign}
		\mu=\ e^{-T/t_{corr}}\times\mu_{-1}+\sqrt{1-e^{-2T/t_{corr}}\ }\times N_\mu \nonumber \\
		\theta=\ e^{-T/t_{corr}}\times\theta_{-1}+\sqrt{1-e^{-2T/t_{corr}}\ }\times N_\theta
	\end{flalign}
	where 
	$T$ is the time elapsed from the last time instant (the last instant when the error was calculated), $\mu_{-1}$ and $\theta_{-1}$ are the amplitude and phase calculated in the last time instant, respectively, and $N_\mu$ and $N_\theta$ are the new uncorrelated samples of the magnitude and angle.
	The parameter $t_{corr}$  
	is used to control the degree of correlation. A smaller $t_{corr}$  causes a quick variation of the error, whereas a larger $t_{corr}$  implies a slow variation of the error. Through our experiments, we derived a $t_{corr}=10$~s.
	The location known at the EGO and included in the messages sent by the EMU is thus obtained in terms of latitude, denoted as $v_\text{lat}$, and longitude, denoted as $v_\text{lon}$, by applying the following equations
	\begin{flalign} \label{eq,GNSS}
	v_\text{lat} = v_\text{lat}^* + \big(\mu\times\sin{(\theta)}\big) \nonumber\\
	v_\text{lon} = v_\text{lon}^* + \big(\mu\times\cos{(\theta)}\big) \end{flalign}
	where $v_\text{lat}^*$ and $v_\text{lon}^*$ are the exact latitude and longitude provided by the \ac{TRS}, respectively.
	
\begin{figure}[]
		\centering
		\includegraphics[scale=0.25,keepaspectratio]{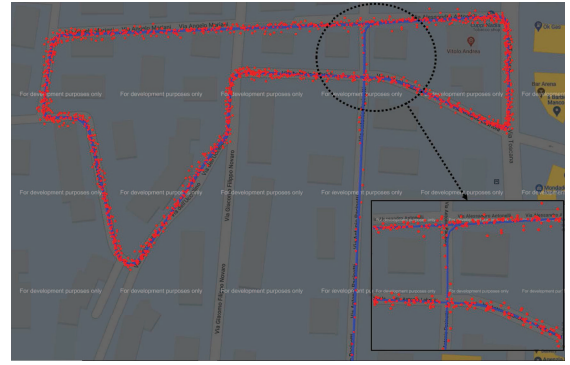}
		\caption{GNSS accuracy error.}
		\label{fig:ACC-GNSS}
	\end{figure}

	Fig. \ref{fig:ACC-GNSS} represents the output of a simulation using the inaccurate \ac{GNSS} positioning model. Specifically, the blue line represents the exact location while the red line provides the estimated location including the positioning error.

\section{Assessing the propagation conditions}\label{sec:Method}

Another aspect that deserved particular attention in our implementation is the derivation of propagation conditions, \red{which means in particular identifying whether the communication link between the two \acp{ITS-S} is in \ac{LOS} or not, before computing the path-loss accordingly.} There are in fact various models proposed in the literature for the calculation of the path-loss in vehicular scenarios, the last one being the one in ETSI TR 103 257-1 \cite{TR103257-1V1.1.1}, summarized in Annex A, and all of them provide different calculations given the \ac{LOS} or \ac{NLOS} conditions, hereafter named \textit{propagation conditions}. Whereas calculating the path loss as a function of the distance is immediate once the propagation conditions are known, identifying if the link between two nodes is obstructed or not might imply more complex calculations. Considering that such an evaluation is required continuously during the simulation due to the mobility of nodes, a huge time and processing effort might be required, possibly compromising the real-time nature of the simulations.

 In this section, we describe the process adopted in our platform to identify the propagation conditions per each link between the \red{EGO} and the other vehicles moving in the scenario. The process is summarized through pseudo code in Algorithm~\ref{Algo1}.

	\begin{algorithm} 
	\footnotesize	
	\caption{Propagation conditions assessment.}
	\LinesNumbered
	\label{Algo1}
	\KwIn{\emph{Vehicle and building positions};} 
	\KwOut{\emph{Propagation conditions};}
	\% Init building positions \\
		\textit{$buildings$} $\leftarrow$ building~positions; \\
			\For{each time step}{
			\% Update vehicle positions \\
		\textit{$e$} $\leftarrow$ ego-vehicle~position;\\
		\textit{$vehicles$} $\leftarrow$ other~vehicle~positions;\\

		\% Reset conditions \\
		$conditions$($\forall v \in vehicles$) = LOS;
		
		\% Select vehicles within $InRangeV$ \\
		\For{$v \in vehicles$}{
			\If{$dist(e, v) < R_{v}$}{$InRangeV \leftarrow v$;}}
		
		\% Select buildings within $InRangeB$ \\
		\For{$b \in buildings$}{
			\If{$dist(e, b) < R_b$}{$InRangeB \leftarrow b$;}}
		
		\% Cycle over the other vehicles \\
		\For{$rv \in InRangeV$}{
		\% Init flag \\
		$flag$ = false;\\
		\% Check NLOS due to buildings \\
			\For {$rb \in InRangeB$}{
				\If{$InBtw(rb, rv, e)$}{
					$flag$ = true \\
					$\textbf{break}$}
				
			}
			
			\If{$flag$}{
			$conditions$(rv) = NLOSb;\\
				$\textbf{break}$} 
			
		\% Check NLOS due to vehicles \\
			\For{$sv \in InRangeV \setminus rv$}{
				\If{$InBtw(sv, rv, e)$}{
					$flag$=true\\
					$\textbf{break}$}
			}
			\If{$flag$}{
				$conditions$(rv) = NLOSv;\\
				$\textbf{break}$}
		\% else LOS conditions: nothing to do \\
		
		
		}
		
	}
	
\end{algorithm}

At the beginning of the simulation, the TRS informs the EMU about the building positions and the static objects (e.g., traffic light or static road works) and provides updated positions of all moving nodes every time interval, hereafter denoted as \textit{step}. Within each step, the simulator should be able to evaluate the quality of all links in order to maintain the real-time nature of the simulation. 
Hence, during each step, the link between the \red{EGO} and each of the other vehicles is categorized as either \ac{LOS}, \ac{NLOSb}, or \ac{NLOSv}.\footnote{Some models do not consider the impact of other vehicles to the path-loss and in such a case only LOS and NLOSb are considered.} A building obstructs the link if any of the walls intersects the segment connecting the \red{EGO} to a target vehicle. Moreover, a third vehicle obstructs the link if its distance, $d_\text{orth}$, from the segment connecting the \red{EGO} and the target vehicle is below a given threshold, with $d_\text{orth}$ calculated as
\begin{flalign}
d_\text{orth}=\frac{\left|mx_\text{sv}-y_\text{sv}-mx_\text{e}+y_\text{e}\right|}{\sqrt{m^2+1}}\;,
\end{flalign}
where $m=\frac{y_\text{rv}-y_\text{e}}{x_\text{rv}-x_\text{e}}$, 
$x_\text{e}$ and $y_\text{e}$ being the \red{EGO} coordinates, 
$x_\text{rv}$, $y_\text{rv}$ the other (target) vehicle coordinates, and $x_\text{sv}$, $y_\text{sv}$ those of the third considered vehicle.

In principle, if $N_\text{v}$ is the number of vehicles in the scenario and $N_\text{b}$ the number of buildings, this operation requires in each simulation step to evaluate $(N_\text{v}-1) \times N_\text{b}$ times if a given building obstructs a link and $(N_\text{v}-1) \times (N_\text{v}-2)$ if a third vehicle obstructs a link, which appears hardly feasible within a single step.

The first and rather obvious observation, in order to reduce the computation effort, is that as soon as \ac{NLOSb} conditions are observed due to one building, there is no need to proceed with the others or with the third vehicles. Similarly, when no building obstructs the link and one of the third vehicles is found to be between the communicating nodes, \ac{NLOSv} conditions are met and there is no reason to proceed with the other third vehicles.
To further reduce the burden of the calculation, two scanning ranges are applied to the \red{EGO}, $R_\text{b}$ and $R_\text{v}$, which limit the considered buildings and other vehicles, respectively, to a maximum distance from the \red{EGO}. The rational for $R_\text{b}$ is that those buildings which are far from the EGO are expected not to relevantly impact on the assessment of NLOSb conditions and it is thus assumed that they can be neglected during the simulation. This assumption is verified through a case study in Section~\ref{sec:Analysis}. The setting of $R_\text{v}$ should instead consider the relevance of the information received by the \red{EGO}; the information received by far vehicles might be not relevant, for example for an intersection collision warning application. In general, $R_\text{v}$ will be not larger than the maximum range for the given settings and $R_\text{b}$ will not be larger than $R_\text{v}$. The buildings and vehicles obtained through this process are hereafter called \textit{relevant objects}. 

In summary, in each simulation step the process detailed in Algorithm~\ref{Algo1} is performed: i) the list of relevant target vehicles $N_\text{v}^\text{(r)}$ (those within $R_\text{v}$) and the list of relevant buildings (those within $R_\text{b}$) are evaluated; ii) a cycle over the  $N_\text{v}^\text{(r)}-1$ vehicles around the \red{EGO} is performed and per each of them, first the \ac{NLOSb} conditions (considering the relevant buildings only) and, in case the \ac{NLOSv} conditions (considering only the relevant vehicles) are verified.

Once the propagation conditions are evaluated, the calculation of the path-loss, the addition of correlated shadowing, and the definition of lost messages are done with marginal addition of computation effort. At the output of this process, the messages that are evaluated to be correctly received by the \red{EGO} (\acp{CAM} in our case study) are obtained; they are then passed to the function in charge to introduce the positioning inaccuracy (described in Section~\ref{sec:GNSS}) and then transmitted through the real ITS-S.

\section{Performance evaluation} \label{sec:Analysis}
\begin{figure}
		\centering
		\includegraphics[scale=0.28,keepaspectratio]{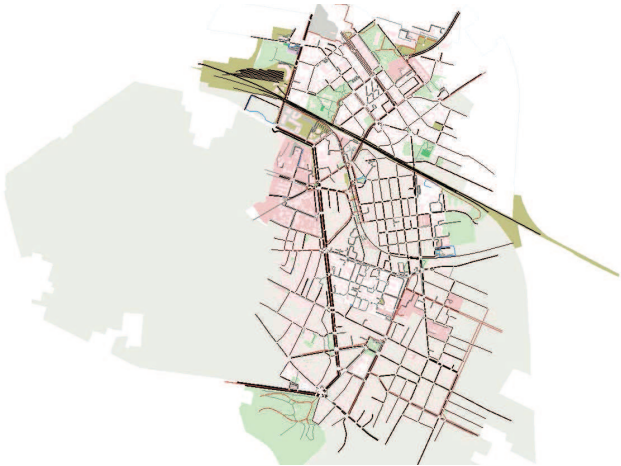}
		\caption{Map of the simulated area.}
		\label{map}
	\end{figure}

To verify the performance of the solutions described in the previous sections, and especially by the proposed propagation conditions evaluation, we considered a portion of the city of Bologna with a maximum point-to-point distance of 3.2~km as shown in Fig.~\ref{map}. 
Vehicles, simulated with \ac{SUMO}, periodically transmit \acp{CAM} using IEEE 802.11p, which exploits a 10 MHz channel in the 5.9 GHz frequency band; transmission power $P_\text{t}= 23$~dBm and sensitivity $P_\text{r}=-82$~dBm are assumed. The urban channel model described in  ETSI TR 103 257-1 \cite{TR103257-1V1.1.1}, reported in Annex A, is adopted with log-normal correlated shadowing characterized by 3~dB standard deviation and 10~m decorrelation distance (as suggested in 3GPP TR 36.885 \cite{3GPPTR36.885V14.0.0}). 
A laptop equipped with a core-i7 processor with 2.59~GHz CPU frequency was used to run the simulation.

\begin{figure}[] 
		\centering
		\includegraphics[scale=0.24]{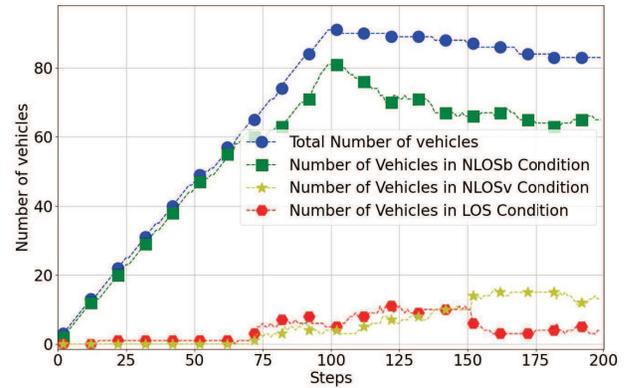}
		\caption{Distribution of traffic.}
		\label{fig:trafic-dis}
	\end{figure}

Fig.~\ref{fig:trafic-dis} illustrates the average number of vehicles at each simulation step in the three different channel conditions (\ac{LOS}, \ac{NLOSb}, and \ac{NLOSv}) as well as the total traffic when the scanning ranges ($R_\text{b}$ and $R_\text{v}$) are set to the 3.2~km, which is the maximum distance of the map and thus is equivalent to infinity. 
The reduction of the scanning ranges, $R_\text{b}$ and $R_\text{v}$, allows to consider a smaller area of the map, thus decreasing the simulation time. 

\begin{figure}[t]
		\begin{center}
		   \subfigure[\label{fig:Acc-NLoS}
		    {\footnotesize Number of vehicles found in \ac{NLOSb} condition.}]
		    {\includegraphics[scale=0.27]{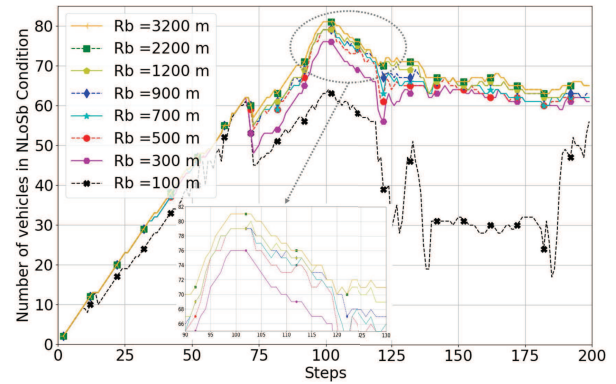}}

	        \subfigure[\label{fig:Acc-Delay}
	        {\footnotesize Simulation delay.}]
	        {\includegraphics[scale=0.26]{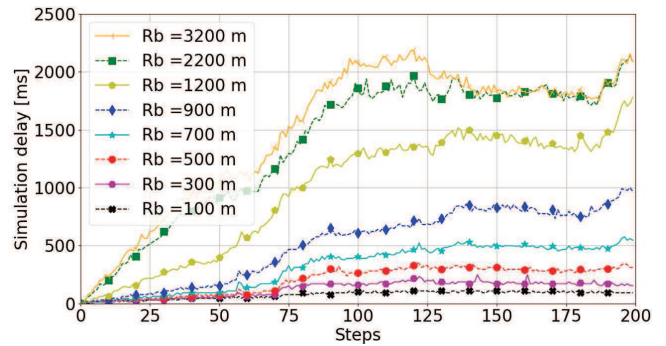}}
	        
        \caption{Simulation performance; varying $R_\text{b}$, with $R_\text{v} = 3.2$~km.}\label{fig:perfAccDelay}
	    \end{center}
    \end{figure}

In Fig.~\ref{fig:perfAccDelay}, the performance of the proposed approximation is measured, with a fixed $R_\text{v}=3.2$~km (i.e., all vehicles in the scenario are considered). 
In particular, the accuracy of the proposed approximation is evaluated in Fig.~\ref{fig:Acc-NLoS} in terms of number of vehicles in \ac{NLOSb} conditions varying the building scanning range $R_b$. The results reveal that by reducing $R_b$ from 3.2~km to 500~m, the number of nodes in \ac{NLOSb} which are not identified as such is around 3\% of the whole nodes; whereas further decreasing this range to 300~m, the same number becomes around 7\%. Moreover, for any value of $R_b$ greater than 900~m, the loss of system accuracy (in terms of missing nodes in \ac{NLOSb} condition) is almost negligible (less than 1\% on average). It should also be remarked that, with $R_b \geq 900$~m, the few nodes not detected in \ac{NLOSb} conditions are all farther than 2~km from the \red{EGO} and thus not really relevant for the performance of the tested applications.

On the other hand, Fig.~\ref{fig:Acc-Delay} illustrates the simulation delay versus simulation steps for the same buildings scanning ranges $R_b$, keeping $R_\text{v}=3.2$~km. \red{The simulation delay is the processing delay introduced by the \ac{EMU} from the instant when it receives the information from the \ac{TRS} to the instant when it forwards the results to the hardware and should be as low as possible to allow the real-time integration of \ac{HiL}}. As observable, when $R_b$ decreases to 500~m and 300~m, the maximum delay decreases from around 2.2~s to 350~ms and less than 200~ms, respectively.

\begin{figure}[]
	\centering
	\includegraphics[scale=0.27]{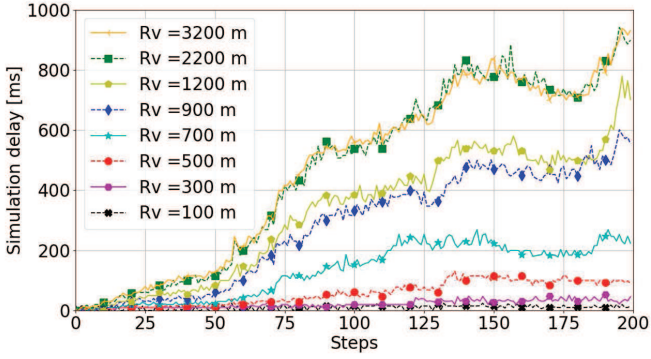}
	\caption{Simulation delay; varying $R_\text{v}$, with $R_\text{b} = 900$~m.}
	\label{fig:Acc-Delay-900}
\end{figure}

The simulation delay is then plotted in Fig.~\ref{fig:Acc-Delay-900} by fixing the $R_\text{b}=900$~m and varying the $R_\text{v}$. The $R_\text{v}$ can be reduced depending on the application's obligation. For example, for some safety applications such as \ac{ICW} that do not need to cover the nodes in a wide area, reducing the $R_v$ to 300~m, the maximum simulation delay would be around 50~ms. In this case, it is not significant to evaluate the accuracy as the range of 900~m for $R_\text{b}$ almost guarantees it.

\begin{figure}[] 
		\centering
		\includegraphics[scale=0.36,keepaspectratio]{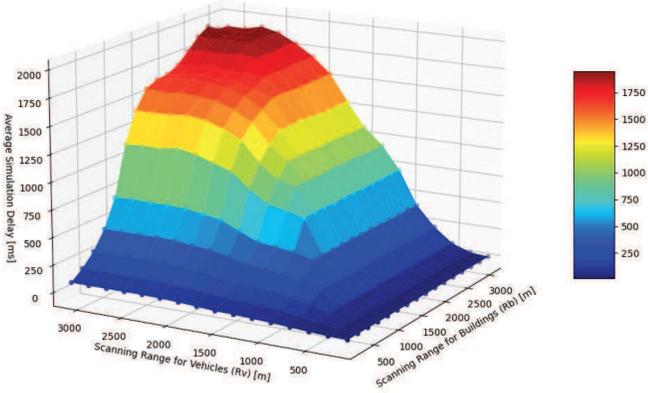}
		\caption{Average simulation delay as a function of the two scanning ranges $R_\text{v}$ and $R_\text{b}$.}
		\label{fig:average-delay}
	\end{figure}

The impact of the approach is finally evaluated in  Fig.~\ref{fig:average-delay} in terms of averaged delay, over the 50 simulation steps with highest traffic, varying both $R_\text{v}$ and $R_\text{b}$. Specifically, Fig.~\ref{fig:average-delay} 
shows that the simulation delay can be reduced down to 18~ms for $R_\text{v}=R_\text{b}= 100$~m and that remains lower than 1~s (i.e., the larger time interval between CAM generations \cite{ETSITS102637-2V1.2.1}) for several combinations of the two scanning ranges. 

	\section{Conclusion and future directions} \label{sec:conclusion}

In this paper, we addressed the topic of real-time \ac{HiL} simulation platforms to validate \ac{V2X} applications in a controlled lab environment. After illustrating the architecture of the validation platform under development, solutions to consider localization accuracy and impact of wireless channel impairments were detailed. The performance of the proposed solutions were derived with reference to a case-study in an urban environment and illustrated that our simulation platform is able to work at run-time with reasonable delay, compatible with most of the safety applications defined in ETSI TS 102 637-2. \red{Currently, small channel load is considered and the access protocols are assumed ideal, with no losses due to collisions; as a next step, models taking into account the current conditions (vehicle density, access technology with modulation and coding scheme, message size, etc.) will be derived to overcome this limitation and still allow real-time processing required when hardware is in the loop.}
The detailed software will be part of a \ac{HiL} platform integrating \ac{ADAS} and connectivity for the validation of \acp{CAV}.

\section*{ACKNOWLEDGMENT}
This activity is part of the project "Design and development of an experimental platform for validation of ADAS and V2X functions for a safe and sustainable mobility", partially funded by Regione Emilia Romagna within regulation "Legge Reg.14/2014 s.m.i."- "POR FESR 2014-2020 e POR FSE 2014-2020: Accordi regionali di insediamento e sviluppo delle imprese".

\section*{Appendix A: path loss model}

The ETSI TR 103 257-1 \cite{TR103257-1V1.1.1} is used here as path-loss model while other channel models can be found in \cite{meinila2009winner, schneider2010large,meinila2010d5}.
	The considered model distinguishes between highway and urban scenarios, as well as \ac{LOS}, \ac{NLOSb}, and \ac{NLOSv} conditions. 
	
		In particular, the following equations model the path loss for the urban environment in LOS and NLOSb
	\begin{align} 
		{P_\text{l}}_\mathrm{LOS}=38.77+16.7 \log_{10} (d_\text{3D})+18.2 \log_{10} (f_\text{c}) \label{Pathloss}\\
		{P_\text{l}}_\mathrm{NLOSb}=36.85+30 \log_{10} (d_\text{3D})+18.9 \log_{10} (f_\text{c}) \label{Pathlossb}
	\end{align}
	where $d_\text{3D}$ is the euclidean transmitter-receiver distance and $f_\text{c}$ is the central frequency in GHz (set to 5.9).
	
	\noindent The attenuation for the \ac{NLOSv} conditions is calculated as

\begin{align} \label{Pathlossv}
{P_\text{l}}_\mathrm{NLOSv} = {{P^\prime}_\text{l}}_\mathrm{NLOSv}+{P_\text{l}}_\mathrm{LOS}	
\end{align}
\small{
\begin{align*} \label{Pathloss1}
	{{P^\prime}_\text{l}}_\mathrm{NLOSv}= 
		\begin{cases}
		6.9+ 20 \log_{10}\sqrt{(\nu-0.1)^2+1}+\\+\nu-0.1 ~~~~~~~~~~~~~~~~~~~~~ \text{for}~\nu>0.7\\
		0 ~~~~~~~~~~~~~~~~~~~~~~~~~~~~~~\text{otherwise.}
		\end{cases}\;,
\end{align*}}
 $\nu=\sqrt2\ \frac{H}{r_\text{f}}$, $H$ is the difference in height between the obstacle and the straight link from transmitter to receiver, $r_\text{f}$ is the first Fresnel zone radius that can be approximated as $r_\text{f}=\sqrt{\frac{\lambda d_\text{1} d_\text{2}}{d_\text{1}+d_\text{2}}}$, $d_\text{1}$ and $d_\text{2}$ are the distances from the blocking vehicle to the transmitter and receiver, respectively, and $\lambda$ is the wavelength. 
	\bibliographystyle{IEEEtran}

	\bibliography{V2}

\end{document}